\documentclass[a4paper,11pt]{article}
\usepackage{enumitem}
\usepackage{float}
\usepackage[margin=2.5cm]{geometry}
\usepackage{graphicx}
\usepackage[labelfont=bf,labelsep=period]{caption}
\usepackage{subcaption}
\usepackage{times}
\usepackage{url}
\usepackage{xspace}
\usepackage{numbertabbing}
\usepackage{subcaption}

\usepackage{amsmath} 
\allowdisplaybreaks[2]          

\usepackage{amssymb} 

\usepackage{amsthm} 

\usepackage[usenames,dvipsnames]{xcolor} 
\usepackage{centernot} 
\usepackage[normalem]{ulem} 

\usepackage{hyperref}

\newcommand\myshade{85}
\colorlet{mylinkcolor}{Maroon}
\colorlet{mycitecolor}{RoyalBlue}
\colorlet{myurlcolor}{Aquamarine}

\hypersetup{
  linkcolor  = mylinkcolor!\myshade!black,
  citecolor  = mycitecolor!\myshade!black,
  urlcolor   = myurlcolor!\myshade!black,
  colorlinks = true,
}

\newtheoremstyle{plain-boldhead}
  {\topsep}
  {\topsep}
  {\itshape}
  {}
  {\bfseries}
  {.}
  { }
  {\thmname{#1}\thmnumber{ #2}\thmnote{ (\bfseries #3)}}
\newtheoremstyle{definition-boldhead}
  {\topsep}
  {\topsep}
  {\normalfont}
  {}
  {\bfseries}
  {.}
  { }
  {\thmname{#1}\thmnumber{ #2}\thmnote{ (\bfseries #3)}}
\theoremstyle{plain-boldhead}
\newtheorem{theorem}{Theorem}

\theoremstyle{definition-boldhead}
\newtheorem{definition}{Definition}

\newtheorem{example}{Example}

\floatstyle{ruled}
\newfloat{algo}{tbhp}{algo}
\floatname{algo}{Algorithm}

\mathchardef\mhyphen="2D

\newcommand{\str}[1]{\textsc{#1}}
\newcommand{\var}[1]{\textit{#1}}
\newcommand{\op}[1]{\textsl{#1}}

\newcommand{\false}{\textsc{false}\xspace}
\newcommand{\true}{\textsc{true}\xspace}

\newcommand{\BN}{\ensuremath{\mathbb{N}}\xspace}

\newcommand{\CA}{\ensuremath{\mathcal{A}}\xspace}

\newcommand{\CP}{\ensuremath{\mathcal{P}}\xspace}

\def \ifempty#1{\def\temp{#1} \ifx\temp\empty }

\newcommand{\minuseq}{\mathrel{{-}{=}}}
\newcommand{\pluseq}{\mathrel{{+}{=}}}
\newcommand\balances{\var{balances}\xspace}
\newcommand\name{\var{name}\xspace}
\newcommand\Symbol{\var{symbol}\xspace}
\newcommand\decimals{\var{decimals}\xspace}
\newcommand\totalSupply{\var{totalSupply}\xspace}
\newcommand\balanceOf{\var{balanceOf}\xspace}
\newcommand\owner{\var{owner}\xspace}

\newcommand\from{\var{from}\xspace}
\newcommand\To{\var{to}\xspace}
\newcommand\Value{\var{value}\xspace}
\newcommand\transferFrom{\var{transferFrom}\xspace}
\newcommand\approve{\var{approve}\xspace}
\newcommand\allowances{\var{allowances}\xspace}
\newcommand\spender{\var{spender}\xspace}
\newcommand\allowance{\var{allowance}\xspace}

\newcommand{\consensusnr}{\mathcal{CN}}

\newcommand{\accountset}{\mathcal{A}}

\newcommand{\etal}{\textit{et~al.}\@}

\newcommand{\kAssetTr}{k\mhyphen AT}

\newcommand{\uniquetransfer}
{\mathtt{U}}

\begin{document}

\title{\bf On the Synchronization Power of Token Smart Contracts}

 \author{Orestis Alpos\\ 
   University of Bern \\
   \url{orestis.alpos@inf.unibe.ch}
   \and
   Christian Cachin\\
   University of Bern \\
   \url{cachin@inf.unibe.ch}
   \and
  Giorgia Azzurra Marson\\
  University of Bern \\
  \url{giorgia.marson@inf.unibe.ch}
  \and
  Luca Zanolini\\
  University of Bern\\
  \url{luca.zanolini@inf.unibe.ch}
 }

\date{}

\maketitle

\begin{abstract}\noindent
Modern blockchains support a variety of distributed applications beyond cryptocurrencies, including \emph{smart contracts}, which let users execute arbitrary code in a distributed and decentralized fashion.
Regardless of their intended application, blockchain platforms implicitly assume consensus for the correct execution of a smart contract, thus requiring that all transactions are totally ordered.
It was only recently recognized that consensus is not necessary to prevent double-spending in a cryptocurrency (Guerraoui~\etal, PODC'19), contrary to common belief.
This result suggests that current implementations may be sacrificing efficiency and scalability because they synchronize transactions much more tightly than actually needed. 

In this work, we study the synchronization requirements of Ethereum's ERC20 token contract, one of the most widely adopted smart contacts. Namely, we model a  smart-contract token as a concurrent object and analyze its consensus number as a measure of synchronization power.
We show that the richer set of methods supported by ERC20 tokens, compared to standard cryptocurrencies, results in strictly stronger synchronization requirements.
More surprisingly, the synchronization power of ERC20 tokens depends on the object's state and can thus be modified by method invocations. To prove this result, we develop a dedicated framework to express how the object's state affects the needed synchronization level.
Our findings indicate that ERC20 tokens, as well as other token standards, are more powerful and versatile than plain cryptocurrencies, and are subject to \emph{dynamic} requirements.
Developing specific synchronization protocols that exploit these dynamic requirements will pave the way towards more robust and scalable blockchain platforms.
\end{abstract}

\section{Introduction}

The rise of cryptocurrencies has motivated the development of distributed applications running over blockchain platforms. These applications go far beyond the concept of a decentralized cryptocurrency, as initially envisioned by Bitcoin~\cite{Nakamoto_bitcoin:a}.
Taking this diversity to the extreme, \emph{smart contracts} enable a blockchain to execute arbitrary programs, in a fully decentralized fashion akin to a ``world computer.''
Introduced by Ethereum~\cite{ethereum}, smart contracts come in many different flavors and are the key element in most blockchain projects today.

Regardless of the type of supported smart contract, blockchain platforms rely on a distributed protocol that orders transactions and emulates a ledger data structure. A transaction may be a simple ``coin transfer'' in a cryptocurrency or a complex method call to a decentralized application. For either use-case, it is widely accepted that the blockchain nodes must execute all transactions in the same order to ensure consistency~\cite{swanson15,DBLP:journals/csur/Schneider90}. That is, to ensure that the emulated ledger is consistent, transactions are sent using protocols that implement \emph{total-order broadcast} or \emph{consensus}. Garay~\etal~\cite{DBLP:conf/eurocrypt/GarayKL15} showed such an equivalence formally for the Bitcoin protocol.
This common theme seems to suggest that total order is also \emph{necessary} for the consistency of blockchains.

However, this folklore intuition is wrong: Recent work by Guerraoui~\etal~\cite{DBLP:conf/podc/GuerraouiKMPS19} shows that consensus is not necessary to avoid double-spending in cryptocurrency applications.
After distilling the essence of a cryptocurrency protocol to the problem of realizing a consistent \emph{asset transfer} (AT), the authors cast the latter as a sequential object in the shared-memory model and prove the AT object has \emph{consensus number}~1 in the wait-free hierarchy~\cite{DBLP:journals/toplas/Herlihy91}. In other words, consensus is not needed at all for emulating the functions of Bitcoin!
The consensus number is a well-established tool to express the synchronization requirements of asynchronous concurrent objects. Informally, it provides an upper bound for the number of processes that can be synchronized using (arbitrarily many) instances of a given object.
For cryptocurrencies modeled after Bitcoin that support shared accounts with up to~$k$ owners, Guerraoui~\etal~\cite{DBLP:conf/podc/GuerraouiKMPS19} introduce a \emph{$k$-shared} asset transfer ($k$-AT) object that has consensus number~$k$, which is as powerful as consensus among its $k$ owners.
Going beyond their theoretical elegance, these results are of great practical interest because they pave the way to consensus-free implementations of cryptocurrencies~\cite{DBLP:conf/dsn/CollinsGKKMPPST20,DBLP:conf/wdag/GuerraouiKMPS19}, with higher efficiency and robustness to network partitions.
In this particular case, for example, only the $k$ owners need to reach consensus for spending from the account, provided they have additional means to publicize this widely in the network.

In this work, we investigate the synchronization power of smart contracts.
We observe that although~$k$-AT would allow to \emph{generically} implement any smart contract among~$k$ processes, it remains open whether this level of synchronization is \emph{necessary} for widely-used blockchain applications. 
We focus our attention on smart contracts for Ethereum, which is by far the most important platform for hosting decentralized applications. Moreover, many other networks have adopted its programming model. We present an abstraction of a token object that captures and generalizes the functionality of an ERC20 contract~\cite{erc20}, which forms the basis for countless applications on Ethereum that hold billions today.
Notice that the $k$-AT abstraction~\cite{DBLP:conf/podc/GuerraouiKMPS19} applies to Bitcoin and its UTXO model of a currency. Ethereum, on the other hand, uses accounts, and ERC20 contracts are considerably more powerful than Bitcoin transactions. The additional features of ERC20 make it possible, for example, to let account owners \emph{conditionally} issue transfers to other users of their choosing. 

Empowering account owners to approve other spenders makes the ERC20 token object strictly more powerful than~$k$-AT. In addition, approval of new spenders can be performed flexibly, at any time and for arbitrary amounts of tokens, achieving a dynamic that has no counterpart in the case of $k$-AT.
Because of these differences, the results established for $k$-AT~\cite{DBLP:conf/podc/GuerraouiKMPS19} cannot be lifted to~ERC20 tokens. 
What crucially distinguishes an ERC20 token object from $k$-shared asset transfer is the increased level of dynamicity, which is reflected in its synchronization requirements.
Namely, the consensus number of~ERC20 tokens depends on the number of approved spenders for the same account, which may change as the account owner enables more spenders.
Based on the observations, we develop a formalism to express that the consensus number of a token object can change over time, depending on the object's state.
More concretely, we prove that there exist specific states from which it is possible to solve consensus among~$k$ processes, for every~$k\leq n$ where~$n$ is the number of accounts defined by the token contract. 
Moreover, these states can be reached by letting any of the account owners approve new spenders.

Establishing the synchronization power of smart contracts is important for understanding the level of synchronization that is required to run decentralized applications in a blockchain network. Not every two users must be synchronized on every aspect of their respective states, this only matters when their actions affect each other. Identifying the level of consensus needed for different applications also paves the way for realizing more efficient blockchain networks, which may exploit more parallelism.

\paragraph{Organization.} After discussing related work in Section~\ref{sec:relatedwork}, in Section~\ref{sec:background} we present relevant notations and background concepts. We then describe ERC20 tokens in Section~\ref{sec:tokens}, and we analyze their synchronization requirements in Section~\ref{sec:results}. In Section~\ref{sec:extension}, we discuss other notable token standards and elaborate on extending our results to these tokens.
Section~\ref{sec:conclusion} concludes our work suggesting future research directions.

\section{Related Work}
\label{sec:relatedwork}

\paragraph{Synchronization requirements and blockchain scalability.}
Guerraoui~\etal~\cite{DBLP:conf/podc/GuerraouiKMPS19} observe that consensus is not necessary to realize a decentralized cryptocurrency. 
They propose a shared-memory abstraction for the asset transfer problem as implemented in Bitcoin~\cite{Nakamoto_bitcoin:a} and show that it requires only a minimal level of synchronization. Specifically, they show that asset transfer has consensus number~1 in Herlihy's wait-free hierarchy~\cite{DBLP:journals/toplas/Herlihy91}.
The approach of analyzing the synchronization requirements of shared objects in terms of consensus number has been used by others. For instance,  Cachin~\etal~\cite{DBLP:conf/dsn/CachinJS12} study the consensus number of various cloud-storage abstractions, and find that a key-value store has the weakest synchronization power (i.e., its consensus number is~1) while a replica object requires the strongest synchronization level (i.e., its consensus number is~$\infty$).

Obviating the need to reach agreement on the exact ordering of transactions opens the door to more scalable solutions than the currently deployed, consensus-based blockchains.
In this context, Collins~\etal~\cite{DBLP:conf/dsn/CollinsGKKMPPST20} present a decentralized payment system based on Byzantine reliable broadcast. Guerraoui~\etal~\cite{DBLP:conf/wdag/GuerraouiKMPS19} generalize the Byzantine reliable broadcast abstraction to the probabilistic setting and propose a protocol which efficiently realizes it, with the goal of replacing the usual quorum-based safety notions with stochastic guarantees for consistency in a distributed network.
While the above-mentioned protocols fulfill the synchronization requirements for implementing asset transfer, and can therefore support plain cryptocurrency applications, they are not sufficient for the implementation of generic smart contracts.

Many other approaches have been explored in order to increase blockchain scalability~\cite{DBLP:conf/fc/CromanDEGJKMSSS16}, most prominently ``on-chain'' proposals such as optimized BFT-based consensus protocols~\cite{DBLP:conf/sosp/GiladHMVZ17,DBLP:journals/corr/abs-1807-04938,DBLP:conf/podc/YinMRGA19}, DAG-based protocols~\cite{IOTA,DBLP:journals/iacr/SompolinskyZ18}, and sharding~\cite{DBLP:conf/ccs/ZamaniM018,DBLP:conf/sp/Kokoris-KogiasJ18,polkadot,shardingeth}, as well as ``off-chain'' solutions such as payment channels~\cite{lightningnetwork,DBLP:conf/sp/DziembowskiEFM19} and sidechains~\cite{sidechains}. 
Even though these alternative approaches have received a lot of attention recently~\cite{DBLP:journals/access/ZhouHZB20}, they have not yet been widely adopted in practice.

\paragraph{Smart contracts and Ethereum tokens.}
Ethereum~\cite{ethereum} is the first open-source cryptocurrency platform supporting smart contracts, providing a decentralized virtual machine for executing arbitrary Turing-complete programs.
The ERC20 standard, introduced by Buterin and Vogelsteller~\cite{erc20}, provides functions for handling tokens over Ethereum, allowing users to transfer various types of transferable goods such as digital and physical assets. It formulates a common interface for fungible tokens and has become the most widely-deployed API for implementing a token functionality, with more than half of the overall Ethereum transactions being~ERC20 token transfers~\cite{DBLP:conf/fc/VictorL19}.

\section{Preliminaries}
\label{sec:background}

\subsection{Shared Memory and Synchronization Power of Shared Objects}
\label{sec:background:sharedmemory}

We begin with presenting well-established concepts from the concurrent computing literature. We mostly follow the standard notations and nomenclature~\cite{DBLP:books/daglib/0025983,DBLP:books/daglib/0030596,DBLP:conf/podc/GuerraouiKMPS19}. 

\paragraph{Concurrent objects.} 

We assume a (finite) set~$\Pi$ of processes that communicate in an asynchronous manner by invoking operations on, and receiving responses from, shared objects.
Processes are sequential, meaning that no process invokes a new operation before completing (i.e., receiving the response from) all previously invoked operations.
We assume a \emph{crash-failure} model: a process may halt prematurely, in which case we say the process has crashed. We say that a process is \emph{faulty} if it crashes during its execution, otherwise we say that it is \emph{correct}.

An \emph{object type} (or simply \emph{object}) defines the functionality of shared-memory programming abstractions providing a finite set of operations. 
We consider \emph{concurrent} objects, namely objects which can be accessed by multiple processes simultaneously and concurrently. 
The specification of these objects can be sequential or not, where ``sequential'' means that all correct behaviors of the object can be described with sequences of invocations and responses (traces).
In this paper, we are only concerned with sequential objects.
We define an object type as a tuple~$T = (Q,q_0,O,R,\Delta)$, where
$Q$ is a set of states,
$q_0\in Q$ is an initial state,   
$O$ is a set of operations, 
$R$ is a set of responses,
and $\Delta\subseteq Q \times \Pi \times O \times Q \times R$ defines the valid state transitions. 
We write $(q,p,o,q',r) \in \Delta$ to denote that process~$p$ invokes operation~$o$ on the object in current state~$q$, and the operation completes by returning response~$r$ and causing the object to enter state~$q'$.

An \emph{implementation} for an object type~$T$ is a distributed algorithm describing, for each process, sufficient steps to realize each of the object's operations in such a way that desired safety and liveness properties are met. 
The strongest liveness condition for object implementations is~\emph{wait-freedom}~\cite{DBLP:journals/toplas/Herlihy91}, requiring that every invocation of any object operation terminates, despite process failures. 

\paragraph{Registers.} The simplest object type is a \emph{register}, which defines a shared-memory functionality providing~$\op{read}$ and~$\op{write}$ operations.
Given a register~$R$, a process can write a value~$v$ into~$R$ by invoking~$R.\op{write}(v)$; upon completion of this operation, the process is given~$\str{true}$ in response.
Similarly, a process can initiate a read operation on~$R$ by invoking~$R.\op{read}()$; the process obtains a value~$R$ stores. 
In the paper, we consider \emph{atomic} registers.
Formally, an atomic register provides 
\emph{termination}, i.e., if a correct process invokes an operation, then the operation eventually completes, and
\emph{validity}, i.e., a read that is not concurrent with a write returns the last value written, while a read that is concurrent with a write returns the last value written or the value concurrently being written. Moreover, an atomic register provides \emph{ordering}, i.e., if a read returns a value $v$ and a subsequent read returns a value $w$, then the write of $w$ does not precede the write of $v$. This property implies that every operation of an atomic register can be thought to occur at a single indivisible point in time, which lies between the invocation and the completion of the operation~\cite{DBLP:books/daglib/0025983}.

\paragraph{Consensus.} Another important object type is \emph{consensus}, which allows a set of processes to agree on a value. A consensus object~$C$ provides a single operation~$\op{propose}$. A process can invoke~$C.\op{propose}(v)$ on input a proposal~$v$ as a candidate value to be agreed upon. Every process can call~$\op{propose}$ with their own proposed value, and only one invocation is permitted (i.e., it is a ``single-shot'' object).
Upon completion, the operation returns a value~$d$, called the decided value.  
Besides wait-freedom (a.k.a.~\emph{termination}), 
we require 
\emph{validity}, i.e., the decided value is the proposal~$v$ of some process, and
\emph{consistency}, i.e., every process returns the same decided value~$d$. 

\paragraph{Asset transfer (AT).}
The asset transfer object was proposed by Guerraoui~\etal~\cite{DBLP:conf/podc/GuerraouiKMPS19} as an abstraction for cryptocurrencies.
Let $\accountset$ be a finite set of accounts, $\lvert \accountset \rvert = n$, and let~$\mu\colon \accountset \to 2^\Pi$ denote the owner map that associates each account~$a\in \accountset$ to the set of processes sharing the account. If $\lvert \mu(a) \rvert = k$, we say that~$a$ is a $k$-shared account.

\begin{definition}[Asset transfer]
  \label{def:asset:transfer}
  The \emph{asset transfer} object associated to~$\accountset$ and~$\mu$, denoted by~$AT = (Q,q_0,O,R,\Delta)$, is defined as follows:
  \begin{itemize}
    \item Set $Q$ contains all \emph{balance} maps, i.e.,  
    \begin{equation}
      Q = \{ \beta\colon \accountset \to \BN \}.
    \end{equation}
    \item The initialization map~$q_0 = \beta_0$ assigns an initial balance to each account.
    \item $O$ contains two operations,  
      $O = \{ \op{transfer}(a_s,a_d,v) : a_s,a_d \in\accountset, v\in\BN \}\cup \{ \op{balanceOf}(a) : a\in\accountset \}$, where
    $\op{transfer}(a_s,a_d,v)$ lets the caller process, say $p$, transfer $v$ tokens from a source account~$a_s$ to a destination account~$a_d$, provided that $p\in \mu(a_s)$, and 
    $\op{balanceOf}(a)$ reads the balance of account~$a$.
    \item $R$ contains the possible responses to operations in~$O$, $R=\{\str{true}, \str{false}\} \cup \mathbb{N}$.
    \item $\Delta$ defines the valid state transitions. 
    Given a state $q = \beta \in Q$, 
    a process $p \in \Pi$ with account $a_p$, 
    an operation $o \in O$, 
    a response $r \in R$,
    and a new state $q'= \beta'\in Q$, 
    we have $(q,p,o,r,q') \in \Delta$ 
    if and only if either of the following conditions holds:
    \begin{itemize}
    \item $o=\op{transfer}(a_s,a_d,v)
    \;\land\; p \in \mu(a_s) 
    \;\land\; \beta(a_s) \ge v 
    \;\land\; \beta'(a_s) = \beta(a_s) - v 
    \;\land\; \beta'(a_d) = \beta(a_d) + v 
    \;\land\; \forall c \in \accountset \setminus \{a_s, a_d\} : \beta'(c) = \beta(c)
    \;\land\; r = \str{true}$;
    \item $o=\op{transfer}(a_s,a_d,v) 
    \;\land\; (\beta(a_s) < v \;\lor\; p \notin \mu(a_s))
    \;\land\; q' = q
    \;\land\; r = \str{false}$;
    \item $o=\op{balanceOf}(a)
    \;\land\; q' = q
    \;\land\; r = \beta(a)$.
  \end{itemize}
\end{itemize}
If the maximum number of processes sharing an account is~$k$, we name the object a $k$-shared asset transfer, and we denote it by~$k$-AT.
\end{definition}

\paragraph{Synchronization power of shared objects.}
The prominent result by Fischer, Lynch, and Paterson~\cite{DBLP:journals/jacm/FischerLP85} establishes the impossibility of wait-free implementing consensus from atomic registers. This means that consensus requires a higher level of synchronization than atomic registers. In fact, the consensus object is \emph{universal}, in the sense that any shared object described by a sequential specification can be wait-free implemented from consensus objects and atomic registers~\cite{DBLP:journals/toplas/Herlihy91}. Therefore, consensus can be used to reason about the synchronization power of all shared objects (which admit a sequential specification) among a number of processes. 
This leads to the central concept of \emph{consensus number} to express the synchronization power of shared objects.

\begin{definition}[Consensus number~\cite{DBLP:journals/toplas/Herlihy91}]
  The \emph{consensus number} associated with an object~$O$ is the largest number~$n$ such that it is possible to wait-free implement a consensus object from atomic registers and objects of type~$O$, in a system of~$n$ processes.
  If there is no largest~$n$, the consensus number is said to be infinite. Given an object~$O$, we denote its consensus number by~$\consensusnr(O)$. 
\end{definition}

The consensus number allows comparing objects based on their synchronization power, thereby establishing a hierarchy among objects---the consensus hierarchy.
In this work, we leverage the concept of consensus number to study the level of synchronization required for popular smart-contracts tokens.

\begin{theorem}[\cite{DBLP:journals/toplas/Herlihy91}]\label{thm:consensusnr:comp}
Let $O$ and $O'$ be two objects such that $\consensusnr(O) = n$ and $\consensusnr(O') > \consensusnr(O)$. Then there is no wait-free implementation of an object of type~$O'$ from objects of type~$O$ and read/write registers in a system of~$n$ processes.
\end{theorem}

\section{Defining ERC20 Tokens as Shared Objects}
\label{sec:tokens}

In this section, we present a smart contract for transferring tokens defined, by the Ethereum Request for Comment (ERC)~20 specification, and propose a corresponding shared-memory abstraction. 

Tokens are blockchain-based assets which can be exchanged across users of a blockchain platform. Ethereum Request for Comment (ERC) 20 defines a standard for the creation of a specific type, dubbed ERC20 token, one of the most widely adopted tokens on Ethereum. 
ERC20 tokens are transferred through dedicated transactions among Ethereum addresses, and are managed by smart contracts. 
For completeness, we reproduce in Appendix~\ref{app:erc-spec} the algorithmic specification defined in the EIP-20 proposal~\cite{erc20}, with minimal notational changes to ease the comparison with the objects defined in this paper.

The definition of a token object we propose is a generalization of an ERC20 token. The reason for slightly deviating from the original specification (as per Algorithm~\ref{alg:erc-20:specification}, Appendix~\ref{app:erc-spec}) is that it represents a more expressive abstraction: it allows us to reason about synchronization requirements of ERC20 tokens as well as comparing them with the asset transfer object.

Let~$\accountset$ be a finite set of accounts. We assume one account per process, $\lvert \Pi\rvert = \lvert \accountset\rvert = n$, and define a bijection $\omega\colon \accountset\to \Pi$ between accounts and processes, i.e., $\omega(a_i) = p_i$ for all $i\in\{1,\dots,n\}$.
We name~$\omega \colon \accountset \to \Pi$ the \emph{owner map} that associates to each account~$a$ the corresponding process~$\omega(a)$ which owns the account.%
\footnote{Although we use a similar formalism as Guerraoui \etal~\cite{DBLP:conf/podc/GuerraouiKMPS19} to define account ownership, we make the restriction to single-owner accounts to meet the Ethereum-token specification. As we will see in later sections, in Ethereum tokens there are no shared accounts, however, a similar concept is enabled by means of dedicated methods.}
To simplify the notation, we use the shorthand~$a_p$ for the account owned by process~$p$, i.e., such that $\omega(a_p) = p$.

Notice that in the case of asset transfer (cf.~Definition~\ref{def:asset:transfer}), \emph{account ownership} captures a slightly different setting than compared to token objects: the former allows for shared ownership while the latter does not. We make this explicit by using different owner maps~$\mu$ and~$\omega$, respectively.
However, as we explain in detail in the next section, ERC20 tokens offer a richer set of operations that, among others, enables a conditioned form of shared ownership.

Using this notation, below we provide a specification for the \emph{ERC20 token} smart contract using the formalism of shared objects (cf.~Section~\ref{sec:background:sharedmemory}).

\begin{definition}[ERC20 token object]
  \label{def:token:object}
  Let~$\accountset$ be a set of accounts and let~$\Pi$ be the set of corresponding owner processes. 
  A \emph{token object}~$T$ associated to~$\accountset$ consists of a tuple $T = (Q,q_0,O,R,\Delta)$, where:
  \begin{description}
    \item[States:] $Q$ contains all \emph{balances} maps and \emph{allowances} maps, i.e., 
    \begin{equation}
    Q = \{ \beta\colon \accountset \to \BN \} \times \{\alpha\colon \accountset\times \Pi\to\BN \}.
    \end{equation}
    Intuitively, for $a\in\accountset$ and~$p\in\Pi$, $\beta(a)$ indicates the \underline{b}alance of account~$a$, and $\alpha(a,p)$ denotes the amount of tokens that process~$p$ is \underline{a}llowed to spend from account~$a$.
    \item[Initial state:] $q_0 = (\beta_0,\alpha_0)$ denotes the pair of initial account balances and allowances.%
    \item[Operations:] $O$ contains the following operations: 
    \begin{align}  
    O &= \{ \op{transfer}(a_d,v) : a_d \in\mathcal{A}, v\in\BN \} \\
    &\cup \{ \op{transferFrom}(a_s,a_d,v) : a_s,a_d\in\mathcal{A}, v\in\BN \}\\  
    &\cup  \{ \op{approve}(p,v) : p\in\Pi, v\in\BN \} \\
      &\cup \{ \op{balanceOf}(a) : a\in\mathcal{A} \} \\
      &\cup \{ \op{allowances}(a,p) : a\in\mathcal{A}, p\in\Pi \}. 
    \end{align}
    Operation~$\op{transfer}(a_d,v)$ lets the caller process, say $p$, transfer $v$ tokens from its account $a_p$ to destination account~$a_d$; similarly, $\op{transferFrom}(a_s,a_d,v)$ lets the caller process transfer~$v$ tokens from source account~$a_s$ to destination account~$a_d$.
    Operation $\op{approve}(p',v)$ lets the caller process~$p$ authorize another process~$p'$ to transfer up to~$v$ tokens from $p$'s account.
    Finally, $\op{balanceOf}(a)$ reads the balance of account~$a$, and $\op{allowances}(a,p)$ reads  the amount of tokens that process~$p$ is authorized to transfer from~$a$.
    \item[Responses:] $R$ contains the possible responses for all operations in~$O$, namely $R=\{\str{true}, \str{false}\} \cup \mathbb{N}$.
    \item[Sequential specification:] $\Delta$ defines the valid state transitions. 
      Given a state $q = (\beta,\alpha) \in Q$, 
      a process $p \in \Pi$ with account $a_p$, 
      an operation $o \in O$, 
      a response $r \in R$,
      and a new state $q'= (\beta',\alpha')\in Q$, 
      we have $(q,p,o,r,q') \in \Delta$ 
      if and only if either of the following conditions holds:
    \begin{itemize}
    \item $o=\op{transfer}(a_d,v) %
    \;\land\; \beta(a_p) \ge v %
    \;\land\; \beta'(a_p) = \beta(a_p) - v %
    \;\land\; \beta'(a_d) = \beta(a_d) + v %
    \;\land\; \forall c \in \mathcal{A} \setminus \{a_p, a_d\} : \beta'(c) = \beta(c) %
    \;\land\; \alpha' \equiv \alpha %
    \;\land\; r = \str{true};$
    \item $o=\op{transfer}(a_d,v) %
    \;\land\; \beta(a_p) < v %
    \;\land\; q' = q %
    \;\land\; r = \str{false}.$
    \item $o = \op{approve}(\bar{p},v) %
    \;\land\; \alpha'(a_p,\bar{p}) = v %
    \;\land\; \alpha'(a,p) = \alpha(a,p)\; \forall (a,p)\neq(a_p,\bar{p}) %
    \;\land\; \beta' \equiv \beta %
    \;\land\; r = \str{true};$
    \item $o=\op{transferFrom}(a_s,a_d,v) %
    \;\land\; \beta(a_s) \ge v %
    \;\land\; \alpha(a_s,p) \ge v %
    \;\land\; \beta'(a_s) = \beta(a_s) - v %
    \;\land\; \alpha'(a_s,p) = \alpha(a_s,p) - v %
    \;\land\; \beta'(a_d) = \beta(a_d) + v %
    \;\land\; \alpha'(a,p) = \alpha(a,p)\; \forall (a,p)\neq(a_s,p) 
    \;\land\; \forall c \in \mathcal{A} \setminus \{a_s,a_d\} : \beta'(c) = \beta(c) %
    \;\land\; r = \str{true};$
    \item $o=\op{transferFrom}(a_s,a_d,v) %
    \;\land\; \left( \beta(a_s) < v \vee \alpha(a_s,p) < v \right) %
    \;\land\; q' = q %
    \;\land\; r = \str{false};$ 
    \item $o=\op{balanceOf}(a) %
    \;\land\; q = q' %
    \;\land\; r = \beta(a);$
    \item $o=\op{totalSupply} %
    \;\land\; q = q' %
    \;\land\; r = \sum_{a \in \mathcal{A}} \beta(a);$
    \item $o=\op{allowances}(a,\bar{p}) %
    \;\land\; q' = q %
    \;\land\; r = \alpha(a,\bar{p}).$    
    \end{itemize}
  \end{description}
\end{definition}
The example below illustrates the various ERC20 token operations and their interplay.
\begin{example}[ERC20 token: sample execution]
\label{ex:erc-exec}
Consider a set of three processes, $\Pi = \{A,B,C\}$ (Alice, Bob, and Charlie), and corresponding accounts, $\accountset = \{a_A,a_B,a_C\}$. Let~Alice be the deployer of an ERC20 token contract, and suppose Alice provides an initial supply of~$10$ tokens, i.e., $\totalSupply = 10$.
According to the ERC20 specification, the token object~$T$ associated to the contract is initialized as follows:
\[
q_0:\quad 
\balances[a_A,a_B,a_C] = [10,0,0], 
\quad\text{and}\quad 
\]
\[
\forall a\in \accountset : \allowances[a][A,B,C] = [0,0,0].
\]
Starting from this initial configuration, let Alice invoke~$\op{transfer}(a_B,3)$, sending~3 tokens to Bob's account. This operation triggers the transfer of~3 tokens from account~$a_A$ to account~$a_B$ and, upon completion, it causes the following state update:
\[
q_1:\quad \balances[a_A,a_B,a_C] \gets [10 - 3, +3, 0].  
\]  
Let now Bob invoke~$\op{approve}(C,5)$, authorizing Charlie to transfer up to~5 tokens from account~$a_B$. Upon completion, this operation causes the following state update:
\[
q_2:\quad \allowances[a_B] \gets [0,0,+5].  
\]
Upon being approved, let Charlie invoke~$\op{transferFrom}(a_B,a_C,5)$ to transfer~5 tokens from Bob's account to his own account. Despite the fact that Charlie's allowance $\allowances[a_B][C] = 5$ would in principle permit such transfer, Bob's balance $\balances[a_B] = 3$ is currently insufficient. Therefore, the operation returns~$\false$, leaving the state unmodified:
\[
q_3 \gets q_2  
\]
Finally, let Charlie invoke~$\op{transferFrom}(a_B,a_A,1)$ to transfer~1 token from account~$a_B$ to Alice's account.
This time, the amount of tokens to be transferred is below the account balance and, upon completion, the operation triggers the following state update:
\[
q_4:\quad 
\balances[a_A,a_B,a_C] \gets [7+1,3-1,0] 
\quad\text{and}\quad
\allowances[a_B]\gets [0,0,5-1].
\] 
\end{example}

\paragraph{Further notation.}
In later sections of the paper, we will make use of the following shortcut notation. 
For every state~$q\in Q$, we write~$T_q$ to denote the token object initialized with state~$q$, i.e., $T = (Q,q,O,R,\Delta)$. 
Similarly, for~$Q'\subseteq Q$ we write~$T_{Q'}$ to indicate that token object is initialized with any state~$q\in Q'$. 
We will also rely on an auxiliary token object~$T|_{Q'}$, which is obtained from~$T$ by restricting the valid state transitions to remain within~$Q'$, i.e., $T|_{Q'} = (Q',q_0,O,R,\Delta')$ where~$q_0\in Q'$ and $\Delta' = \{ (q,p,o,r,q') \in \Delta :  q' \in Q'\}$.
Finally, we note that in the ERC20 standard (cf.~Algorithm~\ref{alg:erc-20:specification}, Appendix~\ref{app:erc-spec}) the state of the smart contract is fully specified by the arrays~$\balances[\,]$ and $\allowances[\,]$. Namely, for all~$a\in \accountset$ and all~$p\in \Pi$, we have $T.\balances[a] = T.\beta(a)$ and $T.\alpha(a,p) = T.\allowances[a][p]$.

\section{Consensus Number of ERC20 Tokens}
\label{sec:results}

In this section, we study the synchronization power of the ERC20 token object by analyzing its consensus number.

\subsection{Overview of the Results}

The consensus number of an ERC20 token object can be expressed in terms of the maximum number of processes that can transfer tokens from the same account.
This number, denoted below by~$k$, depends on the account balances and allowances defined by the object's state~$q = (\beta,\alpha)$, and hence it can change as the state is updated.
In the rest of this section, we therefore analyze the consensus number of a token object~$T$ for various state configurations. 

\paragraph{Approach and challenges.} 
Given the similarities between the ERC20 token object and the $k$-shared asset transfer object, one may think they have the same consensus number. 
Intuitively, the \op{approve} method in ERC20 tokens allows emulating shared accounts by letting every account owner authorize other processes to transfer tokens from its own account.
In fact, there are at least two peculiarities of ERC20 tokens which depart from $k$-shared accounts. 
Firstly, a $k$-shared asset transfer supports at most~$k$ owners per account, where~$k\leq n$ is fixed upfront (because the owner map~$\mu$ in $k$-AT is \emph{static}). This is in contrast with ERC20 tokens, where each account owner can \emph{dynamically} add and remove spenders at any time of the execution, and the number of valid spenders per account is subject to change as the protocol is ongoing. In other words, an ERC20 token object could be loosely seen as a $k$-shared asset transfer with~$k$ changing dynamically.
Secondly, in $k$-AT the owners of a shared account remain owners for the whole lifetime of the object, i.e., they all can transfer tokens from that account as long as the balance is positive. In ERC20 tokens instead, an approved spender remains a valid spender until it consumes the granted allowance or the account owner decides to revoke the spender's allowance (this can be done by resetting the allowance to the default value~0).

These crucial differences show a separation between the $k$-AT object and the ERC20 token object, and suggest that the two objects meet different synchronization requirements. In particular, it is not possible to apply known results and techniques for~$k$-AT to the case of ERC20 tokens.
Moreover, the approval mechanism to add and remove spenders in ERC20 tokens has subtle implications on the object's synchronization power. 

In the rest of this section, we confirm these observations formally and make precise statements about the consensus number of the ERC20 token object.
We now provide a rather informal summary of our results, which we state in full detail and prove in Section~\ref{sec:results:proofs}.
The statements below hold for every~$k\leq n$.

\paragraph{Lower bound.} There exists a set~$S_k$ of states, which we name \emph{synchronization states}, such that for every~$q\in S_k$ it is possible to wait-free implement a consensus object among~$k$ processes using objects of type~$T_q$ (Theorem~\ref{thm:cn:TSk:geq:k}). Formally: 
\begin{equation}
  \label{eq:summary:CN:lowerbound}
  \consensusnr(T_{S_k}) \geq k. 
\end{equation}
To prove this lower bound, we show that a consensus object supporting~$k$ processes reduces to~$T_q$, with~$q\in S_k$, by presenting a wait-free implementation of consensus for $k$ processes from objects of type~$T_q$ and atomic registers. 
  
\paragraph{Upper bound.} The set of states can be partitioned into $[Q_1,\dots,Q_n]$, with $Q = \cup_{k=1}^n Q_k$, so that for every~$q\in Q_k$, at most~$k$ processes can reach consensus using token objects of type $T_q$ (Theorem~\ref{thm:cn:TQk:leq:k}).
Formally: 
\begin{equation}
 \label{eq:summary:CN:upperbound}
  \consensusnr(T_{Q_k}) \leq k.
\end{equation}
Proving the upper bound turns out to be more involved.
We proceed with an indirect argument, showing that the hypothesis $\consensusnr(T_{Q_k}) = k'> k$ leads to a contradiction. Intuitively, the contradiction is reached because no implementation of consensus for~$k'$ processes from~$T_q$, with~$q\in Q_k$, can be wait-free.

\subsection{Technical Results and Proofs}
\label{sec:results:proofs}

Essentially, we show that an ERC20 token represents a dynamic $k$-shared AT object, where~$k$ depends on the current object's state. Specifically, $k$ is the maximum number of valid spenders for the same account. For each~$k\leq n$, where~$n$ is the total number of accounts, there exists a class~$S_k$ of states, the class of~\emph{$k$-synchronization states}, such that $\forall q\in S_k$, it holds $\consensusnr(T_q) \geq k$.
However, we cannot conclude that $\consensusnr(T) = \infty$. We can only say that if a state~$q \in S_n$ is reached, then we can solve consensus among all processes. That is, there exists a state~$q\in S_n \subset Q$ such that~$\consensusnr(T_q) = n$. 
This is weaker than saying that for every state, we can solve consensus among~$n$ processes. In particular, it is not possible to reach such a state~$q\in S_n$ in a wait-free manner, as we prove later.

Let us first define the sets~$S_k$ of synchronization states formally, then we will provide relevant bounds for the consensus number of an ERC20 token object in a synchronization state.

\paragraph{Enabled spenders.}
For every state~$q  = (\beta, \alpha) \in Q$, let $\sigma_q\colon \mathcal{A} \to 2^\Pi$ denote the mapping associating each account~$a$ to its \emph{enabled spenders} according to~$q$, i.e., the set of processes that are enabled to transfer tokens from account~$a$ w.r.t.~balances~$\beta$ and allowances~$\alpha$ specified by state~$q$. Formally,
\begin{equation}
\label{eq:valid:spenders}
 \sigma_q(a) = \{p\in \Pi : p=\omega(a)\ \lor\  \alpha(a, p) > 0 \}
 .
 \end{equation}
Note that we explicitly include the account owner~$\omega(a)$ in the set of enabled spenders for account~$a$. We conventionally assume that an account with zero balance has only its owner as enabled spender, i.e., $\beta(a) = 0 \Longrightarrow \sigma_q(a) = \{\omega(a)\}$.
Indeed, even if there may be some process~$p$, other than the owner, with positive allowance for account~$a$, i.e., $\beta(a) = 0$ and $\alpha(a,p) > 0$, this process would not be able to transfer tokens from~$a$ unless the balance is increased.

\paragraph{State partition.}
Let~$Q_k$, with~$k \leq n$, be the set of states with exactly~$k$ valid spenders from the same account, i.e., 
\begin{equation}
  \label{eq:definition:Qk}
  Q_k = \{q\in Q : \max_{a\in \accountset} |\sigma_q(a)| = k\}.
\end{equation}
Observe that the subsets~$Q_1, \ldots, Q_n$ define a partition.
Intuitively, we would like to say that each subset is associated to a given level of synchronization, defining a hierarchy~$Q_1 \prec \cdots \prec Q_n$ reflecting the synchronization level, where~$Q_k$ corresponds to consensus number~$k$. 
Importantly, the level of synchronization may change as the object's state is updated. In fact, for every~$k$ and for all states~$q\in Q_k$, there exists a valid transition $(q,p,o,r,q')\in \Delta$, such that
  \begin{equation}
    q\in Q_k,
    \quad 
    p = \omega(a),
    \quad 
    o = \op{approve},
    \quad 
    r = \true,
    \quad
    q'\in Q_{k+1},
  \end{equation}
 in the sense that it is possible to reach some state in~$Q_{k+1}$ from~$q\in Q_k$. However, \emph{the only way to do so} is by letting \emph{the owner} of a $k$-spender account~$a$ approve a new spender.

\paragraph{Synchronization states.}
Later in this section, we show how to implement consensus from an ERC20 token object. Intuitively, we leverage an account for which multiple spenders have been approved: we let the spenders engage in a ``race'' where they compete for spending the account's tokens, and the ``winner'' of this competition gets to choose the decided value (in the consensus protocol).
This idea crucially relies on the fact that there is a \emph{unique} winner. To guarantee this, we need to impose an additional requirement on the balance and allowances of the account used in the implementation.
We formally specify such requirement by defining predicate~$\uniquetransfer\colon \accountset \times Q\to \{\true,\false\}$ (ensuring \underline{u}nique transfers) as follows.
Namely, given a state~$q = (\beta,\alpha)$ and an account~$a$, we define: 
\begin{multline}
  \label{eq:unique:transfer}
  \uniquetransfer(a,q)
  \quad\text{if and only if}\quad
  \beta(a)> 0 \quad\land\\
  \left( |\sigma_q(a)| \leq 2
    \quad\lor\quad
  \forall p_i, p_j \in \sigma_q(a)\setminus \{\omega(a)\}\ :\
  \alpha(a,p_i) + \alpha(a,p_j) > \beta(a)\right).  
\end{multline}
We introduce further notation to identify relevant states which will appear in our main results.
For every~$k$ as above, we define $S_k\subset Q_k$ to be the set of states~$q$ with exactly~$k$ valid spenders from the same account~$a$ and such that predicate~$\uniquetransfer$ holds for~$(a,q)$: 
\begin{equation}
  \label{eq:definition:sync:states}
  S_k = \{ q\in Q\ :\ \exists a\in \accountset\ :\ |\sigma_q(a)| = k\ \land\ \ \uniquetransfer(a,q)\}
\end{equation}
We refer to the states in~$S_k$ as \emph{$k$-synchronization states}.
Intuitively, $q\in S_k$ are the states from which we can solve consensus for~$k$ processes, i.e., using an object type~$T_q$, but not for more than~$k$ processes.

\begin{theorem}
  \label{thm:cn:TSk:geq:k}
 For every~$k\leq n$ it holds $\consensusnr(T_{S_k}) \geq k$.
\end{theorem}
\begin{proof}  
We show an implementation of a consensus object~$C$ for~$k$ processes, using an instance of a~$T_q$ object, with $q\in S_k$, and~$k$ atomic registers~$R[1],\dots,R[k]$.
By the hypothesis~$q\in S_k$, at least one account has~$k$ enabled spenders (cf.~\eqref{eq:definition:sync:states}) and satisfies the requirements defined by predicate~$\uniquetransfer$ (defined in~\eqref{eq:unique:transfer}) with respect to~state~$q = (\beta,\alpha)$.
Without loss of generality, let~$a_1\in \accountset$ denote one such account, and let $\sigma_q(a_1) = \{p_1,\dots,p_k\}$ with~$p_1 = \omega(a_1)$. 
Let~$B = \beta(a_1)$ and~$A_j = \alpha(a_1,p_j)$, $j\in \{2,\dots,k\}$, denote the balance of~$a_1$, resp., the allowances of processes~$p_2,\dots,p_k$ for account~$a_1$, w.r.t.~state~$q = (\beta,\alpha)$.
Finally, let~$a_d$ be any account in~$\{a_2,\dots,a_k\}$.
The code for the implementation is shown in Algorithm~\ref{alg:cn_geq_k}, and described below.

\begin{algo*}
  \vbox{
  \small
    \begin{numbertabbing}\reset
    xxxx\=xxxx\=xxxx\=xxxx\=xxxx\=xxxx\=xxxx\kill
  
  \textbf{State}  \label{}\\
  \> $\var{R}[j] \gets \bot, j \in \{1,\dots,k\}$  \label{}\\
  \> An ERC20 object $T$ initialized such that:  \label{}\\ 
  \>\> $T.\balances[a_1] = B$  \label{}\\
  \>\> $T.\allowances[a_1][p_j] = A_j$, $j \in \{2,\dots,k\}$  
  \label{}\\
  \textbf{operation} \( \op{propose}(v) \) // Code for process $p_i$ \label{}\\  
  \> $R[i].\op{write}(v)$  \label{}\\
  \> \textbf{if} $p_i = p_1$ \textbf{then} \label{}\\
  \>\> $T.\op{Transfer}(a_d,B)$ // Transfer full balance \label{}\\ 
  \> \textbf{else} $T.\op{transferFrom}(a_1,a_d,A_i)$  \label{}\\

  \> \textbf{for} $j \in \{2, \dots, k\}$ \textbf{do}  \label{alg:cn_geq_k:line:forloop}\\
  \>\> \textbf{if} $T.\op{allowances}(a_1, p_j) = 0$ \textbf{then}  \label{alg:cn_geq_k:line:ifcond}\\
  \>\>\> \textbf{return} $\var{R}[j].\op{read}()$ \label{}\\
  \> \textbf{return} $\var{R}[1].\op{read}()$ \label{}

  \end{numbertabbing}
  }
  \caption{Wait-free implementation of a consensus object $C$ among $k$ processes in $\{p_1,\dots,p_k\}$ using an ERC20 object $T_q$, with $q\in S_k$, associated to an account set~$\accountset = \{a_1,\dots,a_n\}$.
  }
  \label{alg:cn_geq_k}
  \end{algo*}
  
Briefly, each process~$p_i$ writes its proposed value~$v$ in a register~$R[i]$. Then process~$p_1$ attempts to transfer its whole balance $B$ to account $a_d$, and process $p_i \neq p_1$ invokes operation~$T.\op{transferFrom}$ as an attempt to transfer its whole allowance~$A_i$ from~$a_1$ to~$a_d$. 
Since only one of the \op{transfer} and \op{transferFrom} invocations succeeds (as we prove shortly), we can safely decide the value proposed by the process which triggered the successful transfer.
The intuition is that only the invocation of $\op{transfer}$ by $p_1$ or the first completing invocation of $\op{transferFrom}$ by some process $p_{i^*}$, for $i^* \in \{2,\dots, k \}$, succeeds.
Upon completion of that operation, no other process will be able to issue its own transfer because the balance of $a_1$ will be too low (this is guaranteed by the predicate~$\uniquetransfer$ defined in~\eqref{eq:unique:transfer}).
Moreover, while the allowance of process $p_{i^*}$ will be 0, the rest of the processes will still have positive allowances. Since the allowances can be read by all processes, every process can determine who won the competition and decide the value proposed by the winner.
Therefore, once an operation $\op{propose}$ completes by returning decision value~$v^*$, every other process that invokes $\op{propose}$ also decides~$v^*$.
More precisely, we select the ``winner'' process~$p_{i^*}$ as the one which succeeds in spending its allowance by transferring~$A_{i^*}$ tokens from~$a_1$ to~$a_d$. If none of the processes is found to have zero allowance, then $p_1$ must have been the first that called $\op{propose}$, and thus consumed the whole balance and caused any other calls to $\op{propose}$ to fail.

We now show that the proposed implementation satisfies the \emph{termination}, \emph{validity}, and \emph{agreement} properties of a consensus object (cf.~Section~\ref{sec:background}).
Regarding the termination property, observe that all instructions of operation $C.\op{propose}$ do terminate: writing the proposed value to~$R[i]$ terminates because of the use of an atomic register; the call to \op{transferFrom} terminates because it only involves reading from and writing to registers; the \emph{for} loop is bounded by the number of processes~$k$, and each iteration involves reading the allowance of a process~$p_j$ and potentially reading from the corresponding register~$R[j]$ (termination follows by the properties of register~$R$).
The validity property holds because the decided value is read from one of the registers~$R[i]$ written by process~$p_i$, for $i\in\{1,\dots,k\}$, and the proposal of each process~$p_i$ is written before the~\op{read} operation on that register is invoked (this is enforced by the \emph{if} condition, cf.~line~\ref{alg:cn_geq_k:line:ifcond}).  
 Hence, the decided value must be the proposal of some process~$p_j$, for $j\in\{1,\dots,k\}$. 
As for the consistency property, as we already mentioned, only the first invocation of operation~$\op{transfer}$ or~$\op{transferFrom}$ may succeed.
In the former case, no invocation to~$\op{transferFrom}$ can ever succeed, hence no allowance can be set to~0, hence all processes will return the value proposed by $p_1$.
In the latter case, the allowance of one of the processes~$p_{i^*}$, for $i^*\in\{2,\dots,k\}$ will be decreased from~$A_{i^*}$ to~0, and the \emph{if} condition (cf.~line~\ref{alg:cn_geq_k:line:ifcond}) ensures that only the register written by a process with an allowance of~0 may be read. 
\end{proof}

The previous theorem provides a lower bound for the consensus number of a token object~$T_q$ with initial state~$q\in S_k$. Therefore, so far we can deduce the following inequalities (where the right-most inequality trivially holds):
\begin{equation}\label{eq:cn:Tq:loose:UB}
  k \stackrel{(Thm.\ref{thm:cn:TSk:geq:k})}{\leq} \consensusnr(T_{S_k}) \leq \infty
\end{equation} 
The upper bound in~\eqref{eq:cn:Tq:loose:UB} is a loose one. 
We proceed with establishing a tight upper bound for the consensus number of~$T$. Similarly to the case of the lower bound, we will need to condition our statement on the object's state. 

Observe that starting from the initial state~$q_0$ as defined in the original ERC20 specification---i.e., no process is authorized to issue transfers from accounts they do not own, and all but the contract deployer have zero balances (cf.~Algorithm~\ref{alg:erc-20:specification}, Appendix~\ref{app:erc-spec})---it is possible to reach a state~$q \in S_k$ \emph{as long as} tokens are transferred across accounts, and the owner of an account~$a$ with positive balance approves other~$k-1$ spenders with sufficient allowances.
Therefore, reaching a state in~$S_k$ is conditioned on all these~$k-1$ $\op{approve}$ operations succeeding, and ultimately on the account owner~$p_a$ not failing until then.
Due to the above condition, a \emph{wait-free} implementation of consensus from~$T_{q_0}$ is unachievable. More generally, starting from any state~$q \in Q_k$, it is not possible to wait-free implement consensus among~$k'> k$ processes, as we prove in the following theorem.

\begin{theorem}
  \label{thm:cn:TQk:leq:k}
  For every~$k\leq n$ it holds $\consensusnr(T_{Q_k}) \leq k$.
\end{theorem}
\begin{proof}
We proceed by contradiction and assume a wait-free implementation of consensus for~$k'$ processes using objects of type~$T_{Q_k}$ and atomic registers, where~$k' > k$, hence we show that for any such implementation there exists an infinite sequential execution that leaves it in a bivalent state.
  
Let us first recall some relevant terminology.
A protocol state is \emph{bivalent} if, starting from that state, there exists some execution in which the processes decide~$0$ and some execution in which they decide~$1$.
A protocol state is called \emph{critical} if it is bivalent and any subsequent state, reached by having a process invoke any of the object's methods, is univalent. 
Every wait-free consensus protocol has a critical state~\cite{DBLP:journals/toplas/Herlihy91}. In the following, we denote one such state by~$q_c$.
Further, the invocation which brings the protocol from a critical state to a univalent state is called a \emph{decision step}.

Without loss of generality, let~$p_1, p_2 \in \Pi$ be processes such that the decision step for~$p_1$, denoted by~$o_1$, brings the protocol into a 0-valent state, and the decision step for~$p_2$, denoted by~$o_2$, brings it into a 1-valent state. The rest of the proof is case analysis of the methods which~$p_1$ and~$p_2$ execute in these decision steps.
  
Let us first assume that the decision step for~$p_1$ is to invoke any operation on an atomic register, while the decision step for~$p_2$ is to invoke any operation on a~$T_{Q_k}$ object. Starting from~$q_c$, the sequential execution of~$o_1$ followed by~$o_2$ brings the protocol into a 0-valent state~$q_1$, since $p_1$ took a step first. Instead, the sequential execution of~$o_2$ followed by~$o_1$ brings the protocol into a 1-valent state~$q_2$, since $p_2$ took a step first. However, the states $q_1$ and $q_2$ are identical, because the two operations~$o_1$ and~$o_2$ commute, a contradiction.
  
Let us now assume that at least one of the invocations, say~$o_1$, is on a read-only method. Consider the sequential execution starting from $q_c$, where~$p_1$ executes~$o_1$, then $p_2$ executes~$o_2$, resulting in state~$q_1$, and then~$p_2$ runs alone and terminates. In this execution, $p_2$ must decide~$0$, because $p_1$ took a step first. 
Consider now the execution starting from~$q_c$, where~$p_2$ executes~$o_2$, resulting in state $q_2$, and then~$p_2$ runs alone and terminates. In this execution, $p_2$ decides~$1$. However, the states~$q_1$ and~$q_2$ differ only in the internal values of~$p_1$, since the latter invoked a read-only method, hence they are indistinguishable for~$p_2$. Yet, $p_1$ decides a different value starting from~$q_1$, respectively, $q_2$, a contradiction.
  
According to the commutativity and read-only arguments just described, the decision steps of~$p_1$ and~$p_2$ must operate on the same object and invoke a method that modifies the state of that object~\cite{DBLP:journals/toplas/Herlihy91}.  In the following, we examine all possible combinations for the decision steps, and whenever they commute, or are read-only, we refer to the arguments above to imply a contradiction.
  
Observe that the methods \op{totalSupply}, \op{balanceOf}, and \op{allowance} of the ERC20 token object are read-only, hence we do not examine them further. Moreover, if both~$o_1$ and $o_2$ are \op{approve} invocations, or if one of them is an~\op{approve} invocation and the other is a~\op{transfer} invocation, then~$o_1$ and~$o_2$ commute and a contradiction is reached as shown above. 
We proceed by analyzing the remaining, non-trivial cases.
  
  \noindent\textit{Case 1: both $o_1$ and $o_2$ are  invocations to the \op{transfer} method.} 
  Since \op{transfer} withdraws tokens from the account of the calling process, $o_1$ and~$o_2$ commute except for the case when~$o_1 = \op{transfer}(a_2, x)$, that is, a transfer of~$x$ tokens to the account of~$p_2$, and the  balance of~$p_2$ is not sufficient to execute the transfer~$o_2$ before~$o_1$, that is, $o_2$ returns~$\false$ if executed before~$o_1$. Observe that in this case, $o_2$ is equivalent to a read-only operation, therefore a contradiction is reached as described earlier. 
  (For instance, consider the following two executions: in the first one, $p_2$ executes~$o_1$ and then runs alone, deciding~0; in the second one, operation~$o_2$ is executed first, followed by~$o_1$, hence $p_1$ runs alone and decides~1.) 
  
  \noindent\textit{Case 2: both $o_1$ and $o_2$ are invocations to the \op{transferFrom} method.} 
  These invocations commute, except for the case when they both use the same source account~$a_s$ and the balance of~$a_s$ is only sufficient for one of the two transfers, and both processes are enabled to spend from $a_s$ (without the latter condition the invocation would be equivalent to a read-only operation). 
  Let us focus on this case. Since our implementation solves consensus among~$k'$ processes, and at most~$k$ processes are enabled spenders for the same account, where $k' > k$, there must be (at least) a process~$p_w$ that is not an enabled spender for account~$a_s$---and by definition, $p_w$ cannot be process $p_s = \omega(a_s)$. Assume wlog that the decision step~$o_3$ taken by $p_w$ brings the protocol in a 1-valent state (otherwise swap~$p_1$ for $p_2$ in the following argument). 
  Under this configuration, we will reach a contradiction for any possible method involved in~$o_3$.
  
  Let us begin with the case where~$o_3$ is a~\op{transferFrom} invocation with~$a_s$ as source account, as shown in Figure~\ref{fig:thm:cn:UB:case2}. 
  \begin{figure}
    \centering
    \begin{subfigure}[t]{0.49\textwidth}
        \centering
        \includegraphics[width=0.7\textwidth]{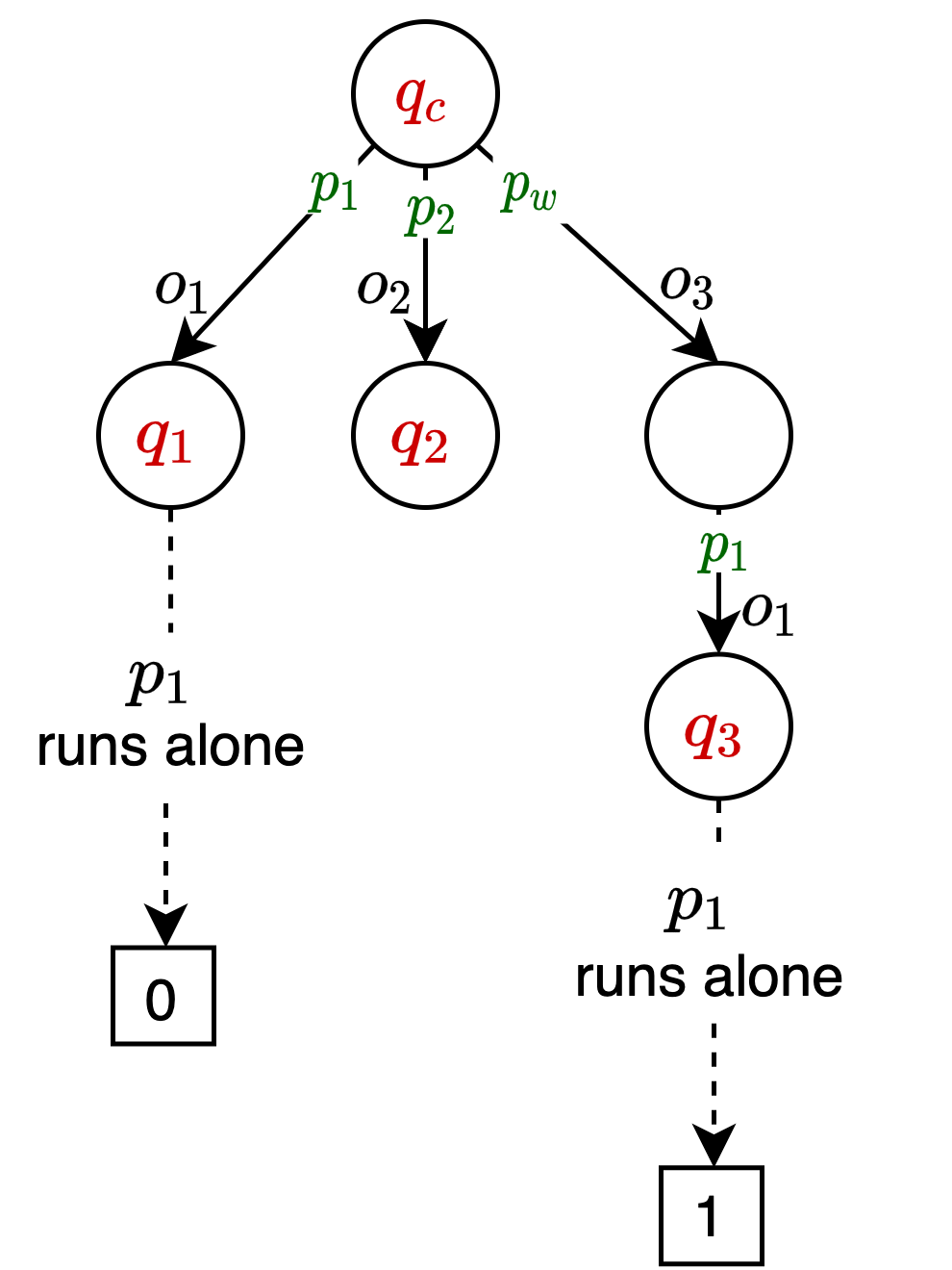}
        \caption{Case 2: both operations~$o_1$ and $o_2$ are invocations to the \op{transferFrom} method.}
        \label{fig:thm:cn:UB:case2}
    \end{subfigure}
    \hfill
    \begin{subfigure}[t]{0.49\textwidth}
        \centering
        \includegraphics[width=0.7\textwidth]{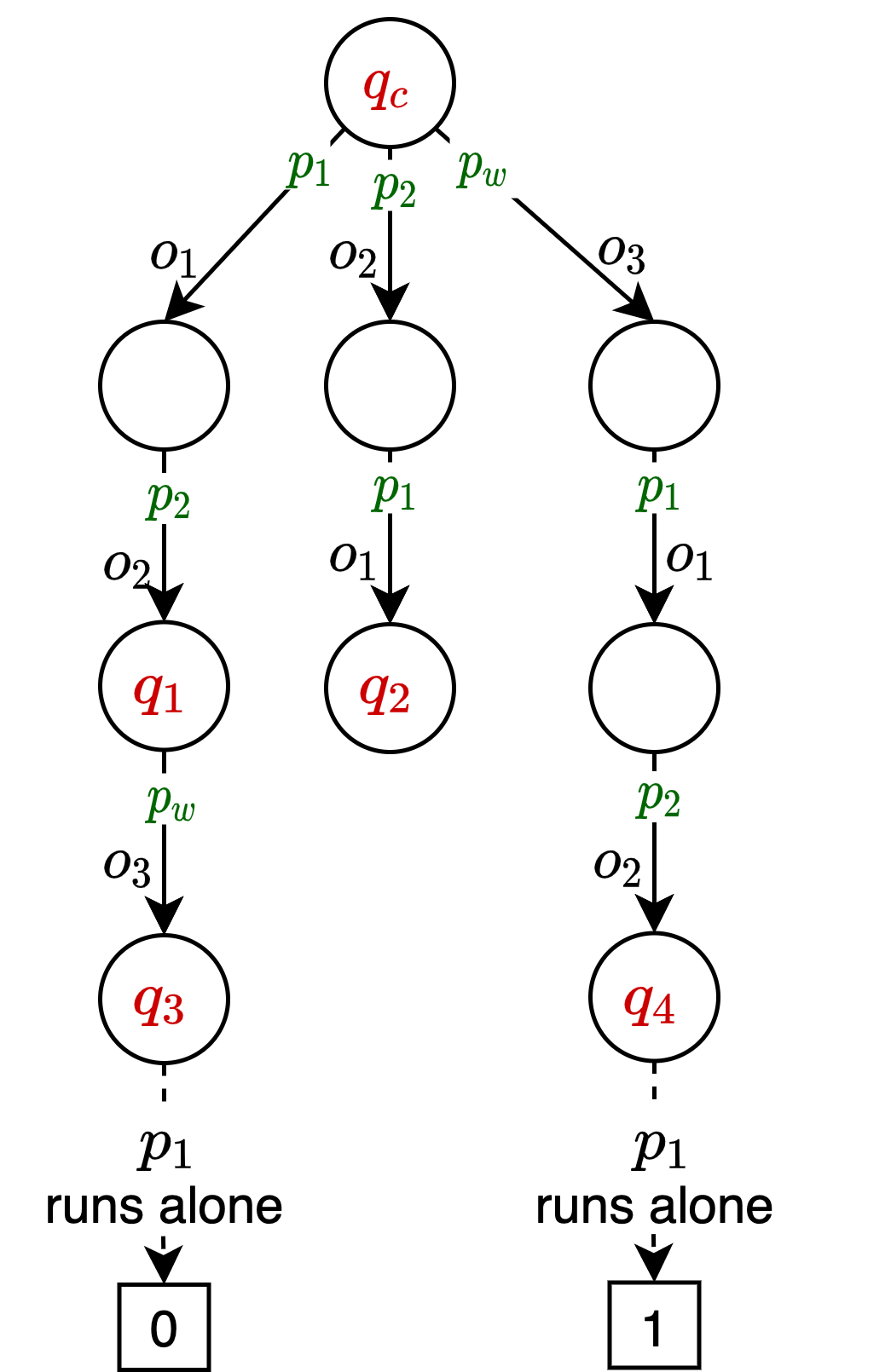}
        \caption{Case 4: operation~$o_1$ is an \op{approve} invocation and~$o_2$ is a \op{transferFrom} invocation.}
        \label{fig:thm:cn:UB:case4}
    \end{subfigure}
  \caption{Possible state transitions from the critical state~$q_c$.}
       \label{fig:thm:cn:UB}
  \end{figure}
  As process~$p_w$ is not enabled for account~$a_s$, operation~$o_3$ returns~$\false$ without modifying the state, thus it is equivalent to a read-only operation.
  Let us now consider the following two executions: 
  process~$p_1$ executes~$o_1$, reaching state $q_1$, and then runs alone, thus deciding 0; 
  process $p_w$ executes~$o_3$, then process~$p_1$ executes~$o_1$ reaching state $q_3$, then process $p_1$ runs alone, thus deciding~1. We have a contradiction, because states~$q_1$ and~$q_3$ are indistinguishable to process~$p_1$.
  
  Moreover, if operation~$o_3$ is a \op{transferFrom} invocation with source account~$a_t$, with~$t \neq s$, then operations~$o_1$ and~$o_3$ commute, and a contradiction is reached with a similar argument as above.
  A similar argument can be applied to all other possible methods, by observing that~$o_3$ is either read-only (\op{totalSupply}, \op{balanceOf}, \op{allowance}), or it commutes with~$o_1$ (\op{approve}, \op{transfer}), because $p_w \neq p_s$.
  
  \noindent\textit{Case 3: operations~$o_1$ and~$o_2$ are a \op{transfer}, respectively, a \op{transferFrom} invocation, or vice versa.} This case is analogous to the previous one. Indeed, if the \op{transferFrom} invocation has a source account other than~$a_1$, then the two invocations commute, while if it has~$a_1$ as source account, the same reasoning as in the previous case, making use of process $p_w$, applies.

\noindent\textit{Case 4: operation~$o_1$ is an \op{approve} invocation and~$o_2$ is a \op{transferFrom} invocation.}
Let us examine the case where~$o_1$ approves process~$p_2$ and~$o_2$ uses~$a_1$ as source account---in all other cases, the two invocations commute. 
We distinguish two cases.

In the first case, assume that~$p_2$ is not already an enabled spender for account~$a_1$. Then operation~$o_2$, if executed before~$o_1$, returns~$\false$ and hence it does not affect the state of~$T_{Q_k}$. Therefore, $o_2$ is equivalent to a read-only operation and a contradiction is reached with the exact same executions as in Case~1 (see above). 
  
In the second case, assume that~$p_2$ is already an enabled spender for account~$a_1$. Then, as depicted in Figure~\ref{fig:thm:cn:UB:case4}, the states~$q_1$ and~$q_2$, reached by the sequential execution of~$o_1$ and then~$o_2$, respectively, by the sequential execution of~$o_2$ and then~$o_1$, are not identical (hence we cannot deduce an immediate contradiction). However, in such case there must be a process~$p_w$ that is not an enabled spender for~$a_1$ and, thus, every possible method invocation~$o_3$ is either read-only or commutes with~$o_1$ and~$o_2$. 
Suppose the decision step taken by~$p_w$ brings the protocol in a~1-valent state (the reasoning for a 0-valent case is analogous). Then the sequential execution of operations~$o_1$, $o_2$, and then~$o_3$ results in a state~$q_3$ from which~$p_1$ decides~0. In contrast, the sequential execution of~$o_3$ followed by~$o_1$ and then~$o_2$ results in a state~$q_4$ from which $p_1$ decides~1. By observing that~$q_3 = q_4$, we reach a contradiction.
\end{proof}

Putting it all together, we have:
\begin{equation}\label{eq:cn:Tq:UB}
  k \stackrel{(Thm.\ref{thm:cn:TSk:geq:k})}{\leq} \consensusnr(T_{S_k}) 
  \quad\text{and}\quad
  \consensusnr(T_{Q_k}) \stackrel{(Thm.\ref{thm:cn:TQk:leq:k})}{\leq} k
  .
\end{equation}
Observing that~$S_k \subseteq Q_k \Longrightarrow \consensusnr(T_{S_k}) \leq \consensusnr(T_{Q_k})$, we can deduce exact synchronization requirements for~$T_q$ when~$q$ is a synchronization state, i.e., $q\in S_k$:
 \begin{equation}\label{eq:cn:Tq:exact}
   k \leq \consensusnr(T_{S_k}) 
  \leq  \consensusnr(T_{Q_k}) \leq k
  \quad\Longrightarrow\quad
  \consensusnr(T_{S_k}) = k
   .
 \end{equation}

Notice that successful completion of specific~$\op{approve}$ operations is \emph{necessary} to reach a synchronization state~$q$ from which we can wait-free implement consensus for arbitrarily many processes. 
Concretely, reaching a state~$q'\in Q_{k'}$ starting from any state~$q\in Q_k$, with~$k' > k$, requires the owner of some account with~$k$ enabled spenders to approve other~$k' - k -1$ spenders.
If such~$\op{approve}$ operations---which change the number of enabled spenders for the same account---were not enabled, then the resulting token object would be no stronger than the $k$-shared asset transfer object.
To make this argument formal, we define an auxiliary token object~$T|_{Q_k}$ by restricting the ERC20 token~$T_q$, with $q\in Q_k$, so that only transitions within~$Q_k$ are permitted. In other words, $T|_{Q_k}$ is a restricted version of~$T$ which does not allow any transition to a state~$q'\in Q\setminus Q_k$. 
The resulting token object reduces to~$k$-shared asset transfer, as we show next.

\begin{theorem}
  \label{thm:reduction:to:kAT}
  For every~$k \leq n$, there exists a wait-free implementation of a token object~$T|_{Q_k}$ from objects of type~$k$-AT and atomic registers.
  In particular, $\consensusnr(T|_{Q_k}) \leq \consensusnr(k\mhyphen AT) = k$.
\end{theorem}
\begin{proof}
  We present a wait-free implementation of object~$T|_{Q_k}$ from a $k$-shared asset transfer object~$\kAssetTr$ (cf.~Definition~\ref{def:asset:transfer}), which is known to have consensus number~$k$~\cite{DBLP:conf/podc/GuerraouiKMPS19}, and atomic registers. The result is implied by the inequality $\consensusnr(T|_{Q_k}) \leq \consensusnr(\kAssetTr)$, which follows from Theorem~\ref{thm:consensusnr:comp}.
  
  Intuitively, for~$q\in Q_k$, we envision every account~$a$ that has~$k$ enabled spenders w.r.t.~$q$ as a $k$-\emph{shared} account, so that we can emulate the methods of object~$T$ by invoking the methods of~$\kAssetTr$ and using atomic registers. Indeed, the token object~$T$ is similar to $k$-shared asset transfer, the crucial difference being that each account owner can approve new spenders for their account and, therefore, can in principle realize~$k'$-shared accounts with $k' > k$. However, the latter operations are disabled for the ``restricted'' object~$T|_{Q_k}$, hence all operations can be simulated by the methods of~$\kAssetTr$ and with registers.  
  
We provide an explicit implementation in~Algorithm~\ref{alg:TQk:from:kAT}.
The balances $\kAssetTr.\beta$ and the owner map $\kAssetTr.\mu$ are initialized in lines~\ref{line:init_kAT_bal} and~\ref{line:init_kAT_mu}, respectively, according to the state $q$.
For each account $a \in \CA$, the algorithm makes use of $n$ atomic registers $R_a[j]$, for $1 \leq j \leq n$, to keep track of the allowance that the owner of $a$ has assigned to process $p_j$. These registers are initialized in line~\ref{line:init_ra} according to the allowance of each account in state $q$.

Recall that the owner map~$\kAssetTr.\mu$ of an asset transfer object is static, i.e., it is defined upon creation of the object. Hence, in order to keep track of dynamically evolving allowances for~$T|_{Q_k}$, we make use of multiple instances of a $k$-AT object: whenever the set of enabled spenders for a given account changes (and as long as the account has no more than $k$ enabled spenders), we create a new instance of the $k$-AT object, with the same balances as the previous instance and an owner map reflecting the updated allowances. In the pseudocode, this is expressed by updating $\kAssetTr.\mu$ in lines~\ref{line:re_first}--\ref{line:re_last}.
  \begin{algo*}
    \vbox{
    \small
      \begin{numbertabbing}\reset
      xxxx\=xxxx\=xxxx\=xxxx\=xxxx\=xxxx\=xxxx\kill
    
    \textbf{State}  \label{}\\
    \> \textbf{for} $a \in \mathcal{A}$ \textbf{do}   \label{}\\
    \>\> $\kAssetTr.\beta[a] \gets \beta(a)$ // Balance of account $a$ in state $q$   \label{line:init_kAT_bal}\\
    \>\> $\kAssetTr.\mu[a] \gets \sigma_q(a)$ // The enabled spenders of account $a$ in state $q$  \label{line:init_kAT_mu}\\
    \> \> \textbf{for} $p_j \in \Pi$ \textbf{do}   \label{}\\
    \> \>\> $R_a[j]\gets \alpha(a, p_j)$ // Allowance of account $a$ to process $p_j$ in state $q$  \label{line:init_ra}\\
    \\
    \textbf{operation} \( \op{transferFrom}(a_s, a_d,\var{value}) \)  // Transfer~$\var{value}$ from source~$a_s$ to destination~$a_d$\label{}\\  
    \> \textbf{if} $R_{a_s}[i] < \var{value}$ \textbf{then}   \label{}\\
    \>\> \textbf{return} \false   \label{}\\
    \> $R_{a_s}[i]  \minuseq \var{value}$   \label{}\\
    \> $\kAssetTr.\op{transfer}(a_s,a_d,\var{value})$   \label{}\\
    \\
     \textbf{operation} \( \op{transfer}(a_d,\var{value}) \)  // Transfer~$\var{value}$ from source~$a_i$ to destination~$a_d$\label{}\\  
    \> \textbf{return} $\kAssetTr.\op{transfer}(a_{p_i},a_d,\var{value})$   \label{}\\ 
    \\
      \textbf{operation} \( \op{balanceOf}(a) \)  // Read balance of~$a$ \label{}\\  
    \> \textbf{return} $\kAssetTr.\op{balanceOf}(a)$   \label{}\\ 
    \\
        \textbf{operation} \( \op{approve}(p_j,\var{value}) \) // Approve spender~$p_j$ for account~$a_i$ \label{}\\        
        \> \textbf{if} $|\{ p_a \} \cup \{p_j \in \Pi : R_a[j] > 0 \}| = k$ \textbf{then}   \label{}\\
        \> \> \textbf{return} $\false$  // Ensure we stay in $Q_k$ \label{}\\
        \> \( \var{oldValue} \gets R_{a_i}[j] \)  \label{}\\
        \> $R_{a_i}[j] \gets \var{value}$ \label{}\\
        \> \textbf{if} $\var{oldValue} = 0$ \textbf{and} $\var{value} > 0$ \textbf{then}   \label{line:re_first}\\
        \> \> \textbf{for} $a \in \mathcal{A}$ \textbf{do}   \label{}\\
        \> \> \> $\kAssetTr.\mu[a] \gets \{ p_a \} \cup \{p_j \in \Pi : R_a[j] > 0 \}$   \label{line:re_last}\\
    \> \textbf{return} $\true$   \label{}\\ 
    \\
        \textbf{operation} \( \op{allowance}(a, p_j) \)  // Read allowance of~$p_j$ for account~$a$ \label{}\\  
    \> \textbf{return} $R_{a}[j]$   \label{}\\ 
     \\
        \textbf{operation} \( \op{totalSupply}() \)  \label{}\\  
    \> \textbf{return} $\sum_{a \in \accountset}{\kAssetTr.\beta[a]}$ \label{}
    
    \end{numbertabbing}
    }
    \caption{Wait-free implementation of a token object $T|_{Q_k}$, with initial state $q = (\beta, \alpha) \in Q_k$, from $k$-shared asset transfer objects $\kAssetTr$. Code for process $p_i$.}
    \label{alg:TQk:from:kAT}
    \end{algo*}
  
  To see why the implementation is correct, observe that no process can transfer more tokens than it has; this is ensured by making use of the~$\kAssetTr$ object. Moreover, for every account~$a$ we keep track of the allowances using registers~$R_a[j]$, for $j\in\{1,\dots,n\}$, and the corresponding balances are managed by the $\kAssetTr$ object. 
  
  Notice that by definition of~$T|_{Q_k}$, all~$\op{approve}$ operations are restricted to transitions within~$Q_k$. Therefore, all valid configurations of account balances and allowances will enable at most~$k$ spenders for the same account. This allows treating such accounts as~$k$-shared, and ultimately enables a correct simulation with the methods of the asset transfer object.   
  The implementation is wait-free because we do not make use of any \emph{for} loop, every method is constructed without the need to wait for other processes to complete their operations and the $\kAssetTr$ object is wait-free.
\end{proof}

\paragraph{ERC20 token \textit{vs} $k$-shared asset transfer.} Informally, Theorem~\ref{thm:cn:TSk:geq:k} and Theorem~\ref{thm:reduction:to:kAT} jointly confirm the intuition that the ERC20 token object is more complex than, yet uncomparable with, the $k$-shared asset transfer object. 
On the one hand, the ERC20 token object is similar to $k$-shared asset transfer, in the sense that $k$-shared accounts can be emulated, to a certain extent, having an account owner approving sufficiently many spenders.
On the other hand, for ERC20 tokens any synchronization level can be reached, in principle, by enabling sufficiently many spenders for the same account, suggesting that ERC20 tokens are strictly more powerful, in terms of synchronization level, than~$k$-AT.
Indeed, while the owners of a shared account must be fixed upfront when the contract is deployed, the enabled spenders for an ERC20 account can change dynamically, as the account owner wishes. Similarly, the amount of tokens that each enabled spender is allowed to transfer from that account is flexibly chosen, and can be modified at any time, by the account owner.
This is in sharp contrast with~$k$-shared asset transfer objects, as the latter has a static consensus number.
Nevertheless, increasing the synchronization power in ERC20 tokens cannot be done in a wait-free manner.

\section{Extension to Other Token Standards}
\label{sec:extension}

In this section, we discuss how to extend our results to other token standards on Ethereum beyond ERC20. As of the time of writing, several token implementations have been proposed within the Ethereum project, ERC20 being the major reference among all. 
Some of these proposals are in a testing phase while others have already reached a final phase and have been adopted~\cite{ERC}.
We overview the proposals that have reached the final stage.
  
The ERC777 token standard aims to solve some problems related to ERC20, while maintaining backward compatibility~\cite{erc777}. It defines new features, some of which are similar to those of ERC20, to interact with the tokens. In particular, it defines $\var{operators}$ to transfer tokens on behalf of another address, similarly to the mechanism enabled by the~$\var{allowances}$ in ERC20,
and \emph{hooks}, to simplify the sending process and to offer a single way for sending tokens to any recipient. One of the main differences compared to ERC20 is the mechanism of allowing processes to manage tokens on behalf of others. In ERC20, the~$\op{approve}$ method lets an account owner~$p$ define an amount of tokens that the approved process~$p'$ is allowed to spend on behalf of~$p$. In contrast, an $\var{operator}$~$p'$ in ERC777 is allowed to spend all the tokens owned by the approving process~$p$. 
Nevertheless, it is immediate to extend our results to ERC777. Specifically, both Algorithms~\ref{alg:cn_geq_k} and~\ref{alg:TQk:from:kAT} can be adapted by replacing the approved spenders with the corresponding operators.
 
The ERC721 standard is inspired by ERC20, however, it provides an interface for \emph{non-fungible} tokens~\cite{erc721}. 
In contrast to standard tokens, all non-fungible tokens are \emph{unique}. 
In ERC721, every token is uniquely determined by an identifier~$\var{tokenId}$ and can be individually transferred using a~$\op{transferFrom}$ method. 
Similarly to ERC20, an account owner~$p$ can approve other processes to spend tokens on its behalf by invoking the~$\op{approve}$ method, specifying the process~$p'$ to be approved and a token identifier~$\var{tokenId}$. 
We do not discuss the other methods specified by ERC721, as they fall outside the scope of this work.
Although ERC721 defines tokens of different nature compared to ERC20 tokens, we notice that our techniques and results can also be applied, with some adjustment, to this standard. 
Concretely, Algorithm~\ref{alg:cn_geq_k} can be adapted so that 
it uses a \emph{specific token}, determined by its identifier~$\var{tokenId}$, which all the participating processes are approved to spend;
the winner of this race can then be determined by invoking~$\op{ownerOf}$ with token identifier~$\var{tokenId}$.
Modifying Algorithm~\ref{alg:TQk:from:kAT} to match the ERC721 specification requires more care.
More generally, implementing an ERC721 token object from~$\kAssetTr$ appears challenging, if not impossible, as each ERC721 token is transferred individually rather than collectively, as is the case with fungible tokens. Instead of relying on $\kAssetTr$, however, a series of~$k$-consensus instances could be used, with each instance associated to an ERC721 token, so that~$k$-consensus can be invoked each time a token is spent. 

ERC1155 defines a smart-contract interface for managing multiple token types. 
In particular, it specifies methods that enable the execution of a number of transactions, possibly on different token types, or involving various source and target accounts, within a single method-call.
While it is plausible that ERC1155 tokens inherit the synchronization requirements of~ERC20 tokens, establishing formal requirements would need an in-depth analysis, based on combinations of accounts, which goes beyond the scope of this work.

Finally, the \emph{Payable Token} standard~ERC1363 follows the~$\op{approve}$ and~$\op{transferFrom}$ paradigm of ERC20 tokens, but adds a layer of indirection. Specifically, it allows processes to specify \emph{arbitrary code}, which is executed upon receiving a token through~$\op{transfer}$, $\op{transferFrom}$, or upon completion of an~$\op{approve}$ operation. 
The possibility of executing an arbitrary contracts precludes establishing exact synchronization requirements \textit{a priori}, as this can be arbitrary.

\section{Conclusion and future directions}
\label{sec:conclusion}

Prior work shows that the asset transfer object, providing the basic functionality of a cryptocurrency, has consensus number~1~\cite{DBLP:conf/podc/GuerraouiKMPS19}. This means that implementing a ``plain'' cryptocurrency such as Bitcoin~\cite{Nakamoto_bitcoin:a} does not require synchronization among processes, and hence the consensus layer of Bitcoin could be replaced by a fully asynchronous dissemination protocol, which does not order transactions.
This important result however does not apply to blockchains with richer smart-contract support such as Ethereum.
In fact, enabling the execution of \emph{arbitrary} smart contracts requires agreement among all blockchain nodes.
Nevertheless, it remains open whether \emph{specific} smart contracts need consensus or not, and more generally, which level of synchronization they require.

In this work, we analyze the synchronization requirements of one such smart contract---the ERC20 token standard of Ethereum---through the lens of wait-free implementations, establishing the consensus number of an associated shared-memory token object.
Our results show that an ERC20 token contract may require different levels of synchronization, depending on its state configurations. In other words, the ERC20 token object has a \emph{dynamic} consensus number:
when initialized according to the standard~\cite{erc20}, its consensus number is~1; however, as soon as an account owner approves other spenders for its account, the consensus number of the object may increase. In fact, there exist executions that modify the state so that the consensus number becomes~$k$, for every~$k$ with~$1\leq k \leq n$.

Our results imply that while executing arbitrary smart contracts requires consensus among \emph{all} processes, synchronizing a dedicated subset of participants is sufficient for realistic applications such as token contracts. 
In the case of ERC20 tokens, consensus indeed only needs to be reached among the largest set~$\sigma_q(a)$ of enabled spenders for the same account~$a$; importantly, the exact synchronization requirements can be readily deduced from the current object's state~$q$ by reading the current balances and allowances.
This insight opens up the possibility to deploy realistic smart contracts, such as ERC20 tokens, on more scalable and performant protocols than consensus-based blockchains. 
Namely, the consistency mechanism could be flexibly adapted, during execution, to require higher or lower coordination among nodes depending on the current state of the smart contract, so that only the minimal synchronization requirements are matched.

We suggest as an interesting open problem to develop distributed protocols meeting the dynamic synchronization requirements of ERC20 tokens. Such protocols could replace the consensus layer of traditional blockchain platforms with a more efficient broadcast method, as shown earlier for asset transfer~\cite{DBLP:conf/dsn/CollinsGKKMPPST20}. This would generally work under asynchrony and yet provide an atomic broadcast functionality among every account owner and its enabled spenders.

\bibliography{references, dblpbibtex}
\bibliographystyle{ieeesort}

\clearpage
\appendix

\section{ERC20 Token Standard}
\label{app:erc-spec}

\begin{algo*}
  \vbox{
  \small
  \begin{numbertabbing}\reset
  xxxx\=xxxx\=xxxx\=xxxx\=xxxx\=xxxx\=xxxx\kill

    \textbf{state} \label{}\\
    \> \textit{const} $d$, the process that deployed the contract, used only in the initialization of \balances  \label{}\\
    \> \textit{const string} \name  \label{}\\
    \> \textit{const string} \Symbol  \label{}\\
    \> \textit{const int} \decimals  \label{}\\
    \> \textit{const int} \totalSupply  \label{}\\
    \> \balances[] \( \subseteq \CA \times \BN \), initially \(\balances[d] =\totalSupply, \balances[i] =0 \text{ for } i \neq d\) \label{}\\
    \> \allowances[][] \( \subseteq \CA \times \CP \times \BN \), initially $\emptyset$  \label{}\\
    \label{} \\

    \textbf{operation} \( \totalSupply() \)  \label{}\\  
        \> \textbf{return}  \totalSupply  \label{}\\
      \label{}\\
    
    \textbf{operation} \( \balanceOf(\owner) \) \label{}\\  
      \> \textbf{return} \( \balances[\owner] \)  \label{}\\
     \label{}\\
    \textbf{operation} \( \op{transfer}(\To, \Value) \) // Code for process~$p_i$ \label{}\\
      \>\textbf{if} \( \balances[p_i] < \Value \)  \textbf{then}\label{}\\
        \>	\>	\textbf{return} \false  \label{}\\
      \> \textbf{else}  \label{}\\
        \>	\> \( \balances[p_i]  \minuseq \Value \)  \label{}\\
         \>	\> \( \balances[\To] \pluseq \Value \)  \label{}\\
         \>	\>	\textbf{return} \true  \label{}\\
     \label{}\\
    \textbf{operation} \(\transferFrom(\from, \To, \Value) \) // Code for process~$p_i$ \label{}\\
       \>\textbf{if} \( \allowances[\from][p_i] < \Value \)  \textbf{then}\label{}\\
        \>	\>	\textbf{return} \false  \label{}\\
      \>\textbf{else if} \( \balances[from] < \Value \)  \textbf{then}\label{}\\
        \>	\>	\textbf{return} \false  \label{}\\
      \> \textbf{else}  \label{}\\
        \> \> \( \allowances[\from][p_i] \minuseq \Value \) \label{}\\
        \>	\> \( \balances[\from] \minuseq \Value \)  \label{}\\
         \>	\> \( \balances[\To] \pluseq \Value \)  \label{}\\
         \>	\>	\textbf{return} \true  \label{}\\
     \label{}\\
    \textbf{operation} \( \approve(\spender, \Value) \) // Code for process~$p_i$	 \label{}\\
    \> \( \allowances[p_i][\spender] \gets \Value \) \label{}\\
    \> \textbf{return} \true  \label{}\\
    \label{} \\
    \textbf{operation} \( \allowance(\owner,\ spender) \) \label{}\\    
      \> \textbf{return} \( \allowances[\owner][\spender] \) \label{} \\
     \label{}
  \end{numbertabbing}
  }
  \caption{Sequential specification of ERC20 functionalities.
  }
  \label{alg:erc-20:specification}
  \end{algo*}

\newpage

\end{document}